\documentclass[
aps,%
pre,%
a4paper,%
final,%
reprint,%
showpacs,%
superscriptaddress,%
groupedaddress,%
floatfix,%
showkeys,%
nofootinbib%
]{revtex4-1}

\usepackage{float}
\usepackage{verbatim}
\usepackage{amsmath,amsfonts,amsthm,amssymb,times,mathrsfs,txfonts}
\usepackage[hidelinks]{hyperref}
\usepackage{bm}
\usepackage{multirow}

\newcommand{\field}[1]{\mathbb{#1}}
\newcommand{\fs}[1]{\mathsf{#1}}

\DeclareMathOperator{\diag}{diag}
\DeclareMathOperator{\Wrons}{\mathscr{W}}

\newcommand{\tp}{\intercal}

\newcommand{\ovl}[1]{\overline{#1}}
\newcommand{\bigO}[1]{\mathop{O}(#1)}

\let\Re\relax
\DeclareMathOperator{\Re}{Re}
\let\Im\relax
\DeclareMathOperator{\Im}{Im}
\newcommand{\vv}[1]{\boldsymbol{#1}}
\newcommand{\vs}[1]{\boldsymbol{#1}}

\newtheorem{defn}{Definition}[section]

\newcommand{\wtilde}[1]{\widetilde{#1}}
\newcommand{\et}{\textit{et~al.}}

\usepackage[pdftex]{graphicx}
\graphicspath{{./figs/}}
\usepackage[caption=false]{subfig}

\begin{document}
\title{Discrete Darboux Transformation for Ablowitz-Ladik Systems Derived from
Numerical Discretization of Zakharov-Shabat Scattering Problem}

\author{Vishal Vaibhav}
\email[]{vishal.vaibhav@gmail.com}
\affiliation{} 
\date{\today}

\begin{abstract}
The numerical discretization of the Zakharov-Shabat Scattering problem using 
integrators based on the implicit Euler method, trapezoidal rule and the
split-Magnus method yield discrete systems that qualify as Ablowitz-Ladik
systems. These discrete systems are important on account
of their layer-peeling property which facilitates the differential approach of
inverse scattering. In this paper, we study the Darboux transformation at the
discrete level by following a recipe that closely resembles the Darboux
transformation in the continuous case. The viability of this transformation for
the computation of multisoliton potentials is investigated and it is found that
irrespective of the order of convergence of the underlying discrete framework,
the numerical scheme thus obtained is of first order with respect to the step
size.
\end{abstract}

\keywords{%
Darboux Transformation, Solitons, Discrete Inverse Scattering
}

\maketitle
\section*{Notations}
\label{sec:notations}
The set of real numbers (integers) is denoted by $\field{R}$ ($\field{Z}$) and 
the set of non-zero positive real numbers (integers) by $\field{R}_+$ 
($\field{Z}_+$). The set of complex numbers are denoted by $\field{C}$,
and, for $\zeta\in\field{C}$, $\Re(\zeta)$ and $\Im(\zeta)$ refer to the real
and the imaginary parts of $\zeta$, respectively. The complex conjugate of 
$\zeta\in\field{C}$ is denoted by $\zeta^*$. The upper-half (lower-half) of $\field{C}$ 
is denoted by $\field{C}_+$ ($\field{C}_-$) and it closure by $\ovl{\field{C}}_+$
($\ovl{\field{C}}_-$). The set $\field{D}=\{z|\,z\in\field{C},\,|z|<1\}$
denotes an open unit disk and $\ovl{\field{D}}$ denotes its 
closure. The set $\field{T}=\{z|\,z\in\field{C},\,|z|=1\}$ denotes the unit 
circle. The Pauli's spin matrices are denoted by, 
$\sigma_j,\,j=1,2,3$, which are defined as
\[
\sigma_1=\begin{pmatrix}
0 &  1\\
1 &  0
\end{pmatrix},\quad
\sigma_2=\begin{pmatrix}
0 &  -i\\
i &  0
\end{pmatrix},\quad
\sigma_3=\begin{pmatrix}
1 &  0\\
0 & -1
\end{pmatrix},
\]
where $i=\sqrt{-1}$. For uniformity of notations, we denote
$\sigma_0=\text{diag}(1,1)$. Matrix transposition is denoted by $(\cdot)^\tp$
and $I$ denotes the identity matrix. For any two vectors 
$\vv{u},\vv{v}\in\field{C}^2$, $\Wrons(\vv{u},\vv{v})\equiv (u_1v_2-u_2v_1)$ 
denotes the Wronskian of the two vectors and $[A,B]$ stands for the commutator 
of two matrices $A$ and $B$.

\section{Introduction}\label{sec:discrete-system}
The main focus of this article is to discuss the special cases of the 
Ablowitz-Ladik (AL) systems that arise as a result of numerical discretization
of the Zakharov-Shabat (ZS) scattering problem~\cite{ZS1972} which forms the
starting point for the definition of what is know the \emph{nonlinear Fourier
transform}~\cite{AKNS1974}. The most general AL 
system~\cite{AL1975JMP} can be stated as: 
define $\vv{v}(n;z)=(v_{1}(n;z),v_{2}(n;z))^{\tp}$, then
\begin{multline}\label{eq:AL-general}
\vv{v}(n+1;z)-\begin{pmatrix}
z^{-1} & 0\\
0& z
\end{pmatrix}\vv{v}(n;z)\\
=\begin{pmatrix}
0 & G(n)\\
H(n)& 0
\end{pmatrix}\vv{v}(n;z)
+\begin{pmatrix}
0 & S(n)\\
T(n)& 0
\end{pmatrix}\vv{v}(n+1;z)
\end{multline}
where $G(n)$, $H(n)$, $S(n)$ and $T(n)$ are certain discrete potentials with $z$
as the discrete spectral parameter. It is noteworthy that in the original work of 
Ablowitz~\et~\cite{AL1975JMP,AL1976JMP}, the general form of the AL system does not seem 
to follow from any quadrature scheme for ODEs. However, certain special cases of the 
AL system can be obtained as a result of applying exponential integrators to the
ZS problem. These special cases fall under the following two categories.
\begin{enumerate}
\item[AL$_1$:] In the transfer matrix formalism, this special case of the AL system
can be stated as
\begin{equation}\label{eq:AL1}
\begin{split}
\vv{v}(n+1;z)&=
\frac{1}{\Delta(n)}
\begin{pmatrix}
1 & S(n)\\
T(n)& 1
\end{pmatrix}
\begin{pmatrix}
z^{-1} & 0\\
0 & z
\end{pmatrix}\vv{v}(n;z)\\
&=z^{-1}M(n+1;z^2)\vv{v}(n;z),
\end{split}
\end{equation}
where $\Delta(n)$ depends only on $S(n)$ with $T(n)=-S^*(n)$.
\item[AL$_2$:] In the formalism adopted above, this special case of the AL system
can be stated as
\begin{equation}\label{eq:AL2}
\begin{split}
\vv{v}(n+1;z)&=
\frac{1}{\Delta(n)}
\begin{pmatrix}
z^{-1} & G(n)\\
H(n)& z
\end{pmatrix}\vv{v}(n;z),
\end{split}
\end{equation}
where $\Delta(n)$ depends only on $G(n)$ with $H(n)=-G^*(n)$. This AL system can
be reduced to the first kind by employing the following transformation
\begin{equation}
\vv{v}(n;z)=
\begin{pmatrix}
1  & 0\\
0  & z
\end{pmatrix}\vv{w}(n;z),
\end{equation} 
so that
\begin{equation}
\begin{split}
\vv{w}(n+1;z)&=
\frac{1}{\Delta(n)}
\begin{pmatrix}
1 & G(n)\\
H(n)& 1
\end{pmatrix}
\begin{pmatrix}
z^{-1} & 0\\
0      & z
\end{pmatrix}\vv{w}(n;z),
\end{split}
\end{equation}
\end{enumerate}
Based on the discussion above, it suffices to just consider the system AL$_1$.
In order to treat the system AL$_2$, we first reduce it to AL$_1$ and adapt the
results accordingly. With regard to the discrete Darboux transformation, the AL
system has been studied by a number of authors and it is simply impossible to
survey them here. In particular, the results obtained in this paper can 
also be derived using the procedure described 
by Rourke~\cite{R2004} or Geng~\cite{GENG198921} (this list is by 
no means exhaustive). However, let us remark that the recipe
provided by Geng seems to evaluate the Jost solutions in the region of the
complex plane where it is not analytically continued.

Now turning our attention to the AL problems described above, let us define
the quantity $\Delta(n)$. To this end, let us define
\begin{equation}
C(n+1)=\det[M(n+1;z^2)]=\frac{1-S(n)T(n)}{\Delta^2(n)}.
\end{equation}
It turns out that the AL systems considered in this article tend to 
have either $\Delta(n)=1-S(n)T(n)$ or $\Delta(n)=\sqrt{1-S(n)T(n)}$ so 
that either $C(n+1)=[1-S(n)T(n)]^{-1}$ or $C(n+1)=1$, respectively. The spectral 
norm of the transfer matrix is given by
\begin{equation}
\|M(n+1;z^2)\|_s=\frac{1}{\Delta(n)}\sqrt{1-S(n)T(n)},\quad z\in\ovl{\field{D}},
\end{equation}
which implies $\|M(n+1;z^2)\|_s\leq1$ or $\|M(n+1;z^2)\|_s=1$, respectively.
Either of these situations indicate the stability of the recurrence
relation in~\eqref{eq:AL1} and~\eqref{eq:AL2}.

In the following, we summarize three of the well-known numerical methods for the ZS problem,
namely, the \emph{implicit Euler} method (also known as the \emph{backward differentiation 
formula} of order one), the \emph{split-Magnus} method
and the \emph{trapezoidal rule} of integration. It is important to emphasize
that the manner in which these methods are applied to the ZS problem as discussed 
by Vaibhav~\cite{V2017INFT1} yield what are known as \emph{exponential
integrators} based on an integrating factor~\cite{CM2002}. The first one
leads to a discrete system with a first order of convergence while the latter
two lead to that with a second order of convergence~\cite{V2017INFT1}. Further,
these systems are unique in that they satisfy the \emph{layer-peeling} property
and that they are amenable to FFT-based fast
polynomial arithmetic which makes them an extremely useful tool for developing
fast direct/inverse nonlinear Fourier transform
algorithms~\cite{V2017INFT1,V2017BL,V2018LPT} within the differential approach of
inverse scattering~\cite{BLK1985,BK1987}.

In order to discuss the discretization 
schemes, we take an equispaced grid defined by $t_n= nh,\,\,n\in\field{Z}$ where $h$ is 
the grid spacing. We also consider a method which employs a staggered grid
configuration defined by $t_{n+1/2}=t_n+h/2$ for sampling the potential. Further, it turns out that the
discrete spectral parameter in these problems can be defined as 
$z=e^{i\zeta h}$. For the sake of convenience, we also introduce
\begin{equation}
\Lambda(z) = 
\begin{pmatrix}
1 & 0\\
0 & z
\end{pmatrix}.
\end{equation}
After the introduction of the discrete systems, the exposition is broadly divided in 
two parts. The first part (Sec.~\ref{sec:DDT}) develops the discrete Darboux transformation for each
of the aforementioned discrete systems and the second part (Sec.~\ref{sec:num-test}), which concludes this
paper, describes a numerical experiment to verify the claims made.
\subsection{Implicit Euler method}
The Zakharov-Shabat scattering 
problem~\cite{ZS1972} can be stated as follows:
Let $\zeta\in\field{R}$ and $\vv{v}=(v_1,v_2)^{\tp}\in\field{C}^2$, then 
\begin{equation}\label{eq:zs-prob}
\vv{v}_t = -i\zeta\sigma_3\vv{v}+U\vv{v},\\
\end{equation}
where
\begin{equation}
U=\begin{pmatrix}
0& q(t)\\
r(t)&0
\end{pmatrix},\quad r(t)=-q^*(t),
\end{equation}
is identified as the \emph{scattering potential}. We begin with the 
transformation $\tilde{\vv{v}}=e^{i\sigma_3\zeta t}\vv{v}$
so that~\eqref{eq:zs-prob} becomes
\begin{equation}\label{eq:exp-int}
\begin{split}
\tilde{\vv{v}}_t&=\wtilde{U}\tilde{\vv{v}},\\
\wtilde{U}&=e^{i\sigma_3\zeta t}Ue^{-i\sigma_3\zeta t}
=\begin{pmatrix}
0 & qe^{2i\zeta t}\\
re^{-2i\zeta t} & 0
\end{pmatrix}.
\end{split}
\end{equation}
Setting $Q(n)=hq(t_n)$, $R(n)=-Q^*(n)$ and 
\begin{equation}\label{eq:Theta-bdf1}
\Theta(n)=[1-Q(n)R(n)]>0,
\end{equation}
the discretization of~\eqref{eq:zs-prob} using the implicit Euler method reads as 
\begin{equation}\label{eq:scatter-BDF1}
\begin{split}
\vv{v}({n+1};z)&=\frac{z^{-1}}{\Theta(n+1)}
\begin{pmatrix}
1& Q({n+1})\\
R({n+1})& 1
\end{pmatrix}\Lambda(z^2)\vv{v}(n;z)\\
&=z^{-1}M(n+1;z^2)\vv{v}(n;z),
\end{split}
\end{equation}
where we have used the convention that $\vv{v}(n;z)$ approximates
$\vv{v}(t_n;\zeta)$. It is straightforward to verify that the recurrence 
relation has a bounded solution for all $z\in\ovl{\field{D}}$ on account of the fact that 
$\|M(n;z^2)\|_s=\Theta^{-1/2}(n)\leq1$ for all $z\in\ovl{\field{D}}$.

\subsection{Split-Magnus method}
Unlike the implicit Euler method, the split-Magnus method employs a staggered grid defined by 
$t_{n+1/2}=t_n+h/2$ to sample the potential. Labeling the samples of the potential
accordingly, this method can be stated as
\begin{equation}
\begin{split}
\vv{v}(n+1;z)&=\frac{1}{\Theta^{1/2}\left(n+\frac{1}{2}\right)}
\begin{pmatrix}
z^{-1}& Q\left(n+\frac{1}{2}\right)\\
R\left(n+\frac{1}{2}\right) & z\\
\end{pmatrix}\vv{v}(n;z),
\end{split}
\end{equation}
where 
\begin{equation}
\Theta\left(n+\tfrac{1}{2}\right)=\left[1-Q\left(n+\tfrac{1}{2}\right)
R\left(n+\tfrac{1}{2}\right)\right]>0. 
\end{equation}
By employing the transformation 
\begin{equation}
\vv{w}(n;z)=\Lambda(z^{-1})\vv{v}(n;z),
\end{equation} 
we obtain
\begin{equation}
\begin{split}
\vv{w}(n+1;z)&=\frac{z^{-1}}{\Theta^{1/2}\left(n+\frac{1}{2}\right)}
\begin{pmatrix}
1& Q\left(n+\frac{1}{2}\right)\\
R\left(n+\frac{1}{2}\right)& 1
\end{pmatrix}\Lambda(z^2)\vv{w}(n;z)\\
&=z^{-1}M(n+1;z^2)\vv{w}(n;z).
\end{split}
\end{equation}
Again, it is straightforward to verify that the 
recurrence relation has a bounded solution for all 
$z\in\ovl{\field{D}}$ on account of the fact that 
$\|M(n;z^2)\|_s=1$ for all $z\in\ovl{\field{D}}$.

\subsection{Trapezoidal rule}
Applying the trapezoidal rule to the transformed ZS problem in~\eqref{eq:exp-int}, we
obtain ~\cite{V2017INFT1,V2017BL}
\begin{multline}\label{eq:scatter-TR}
\vv{v}(n+1;z)=\frac{z^{-1}}{\Theta(n+1)}\times\\
\begin{pmatrix}
1&Q(n+1)\\
R(n+1)& 1
\end{pmatrix}
\Lambda(z^2)
\begin{pmatrix}
1&Q(n)\\
R(n)& 1
\end{pmatrix}
\vv{v}(n;z),
\end{multline}
where $Q(n)=(h/2)q(t_n)$, $R(n)=-Q(n)$ and $\Theta(n)$ is defined 
by~\eqref{eq:Theta-bdf1}. Putting
\begin{equation}
\vv{w}(n;z)=
\begin{pmatrix}
1&Q(n)\\
R(n)& 1
\end{pmatrix}
\vv{v}(n;z),
\end{equation}
we have
\begin{multline}\label{eq:TM-TR}
\vv{w}(n+1;z)=z^{-1}\frac{[2-\Theta(n+1)]}{\Theta(n+1)}\times\\
\begin{pmatrix}
1&G(n+1)\\
H(n+1)& 1
\end{pmatrix}\Lambda(z^2)
\vv{w}(n;z)\\
=z^{-1}M(n+1;z^2)\vv{w}(n;z),
\end{multline}
where
\begin{equation}
G(n)=\frac{2Q(n)}{2-\Theta(n)},\quad H(n)=-G^*(n).
\end{equation}
The transformed system is now similar to the split-Magnus method. It is straightforward to show that 
$\|M(n;z^2)\|_s=1$ for all $z\in\ovl{\field{D}}$ so that this recurrence
relation is stable. 
\section{Discrete Darboux Transformation}
\label{sec:DDT}
Let us introduce the following definition for convenience:
\begin{defn}[Para-conjugate] For any scalar 
valued complex function, ${f}(z)$, we define $\ovl{f}(z)= f^*(1/z^*)$. For any vector 
valued complex function, $\vv{f}(z)=(f_1(z),f_2(z))^{\tp}$, we define 
\[
\ovl{\vv{f}}(z)= i\sigma_2\vv{f}^*(1/z^*)=
\begin{pmatrix}
\ovl{f}_2(z),\\
-\ovl{f}_1(z)
\end{pmatrix}.
\] 
For a matrix valued function, $M(z)$, we define 
\[
\ovl{M}(z)=i\sigma_2M^*(1/z^*)(i\sigma_2)^{\dagger}
=\sigma_2M^*(1/z^*)\sigma_2,
\] 
so that the operation $\ovl{(\cdot)}$ is distributive over matrix-vector and matrix-matrix
products.
\end{defn}
This definition uses the following identity for a $2\times2$ matrix:
\begin{equation}
\sigma_2
\begin{pmatrix}
a_{11} & a_{12}\\
a_{21} & a_{22}
\end{pmatrix}\sigma_2=
\begin{pmatrix}
a_{22} & -a_{21}\\
-a_{12} &  a_{11}
\end{pmatrix}.
\end{equation}

The solution of the discrete scattering problem consists in computing the so 
called \emph{Jost solution} defined as follows: The Jost solution of the first kind is
denoted by $\vs{\Psi}(n;z)$ for $z\in\field{T}$, which satisfies the asymptotic boundary
condition 
\begin{equation}
\vs{\Psi}(n;z)\rightarrow
\begin{pmatrix}
0\\
1
\end{pmatrix}z^{n},
\end{equation}
as $n\rightarrow+\infty$. The Jost solution of the second kind is denoted by
$\vs{\Phi}(n;z)$ for $z\in\field{T}$, which satisfies the asymptotic boundary condition
\begin{equation}
\vs{\Phi}(n;z)\rightarrow
\begin{pmatrix}
1\\
0
\end{pmatrix}z^{-n},
\end{equation}
as $n\rightarrow-\infty$. These Jost solutions can be shown to be analytic~\cite{AL1975JMP,AL1976JMP} on
the unit circle ($\field{T}$) under suitable decay condition on $Q(n)$. They also admit 
of analytic continuation into the open unit disk ($\field{D}$). Further, it is also
possible to define a second set of linearly independent Jost solutions which are
analytic outside $\field{D}$:
\begin{equation}
\begin{split}
\ovl{\vs{\Psi}}(n;z)&=i\sigma_2{\vs{\Psi}^*}\left(n;\frac{1}{z^*}\right),\\
\ovl{\vs{\Phi}}(n;z)&=i\sigma_2{\vs{\Phi}^*}\left(n;\frac{1}{z^*}\right).
\end{split}
\end{equation}
with the asymptotic behavior given by
\begin{equation}
\begin{split}
&\text{as}\,\,n\rightarrow+\infty,\quad
\ovl{\vs{{\Psi}}}(n)\rightarrow
\begin{pmatrix}
1\\
0
\end{pmatrix}z^{-n};\\
&
\text{as}\,\,n\rightarrow-\infty,\quad
\ovl{\vs{{\Phi}}}(n)\rightarrow
\begin{pmatrix}
0\\
-1
\end{pmatrix}z^{n}.
\end{split}
\end{equation}
For $z\in\field{T}$, the discrete scattering coefficients, $A(z)$ and $B(z)$, as defined by
\begin{equation*}
\begin{split}
\vs{\Phi}(n;z)&=A(z)\ovl{\vs{\Psi}}(n;z)+B(z)\vs{\Psi}(n;z),\\
\ovl{\vs{\Phi}}(n;z)&=-\ovl{A}(z)\vs{\Psi}(n;z)+\ovl{B}(z)\ovl{\vs{\Psi}}(n;z),
\end{split}
\end{equation*}
so that
\begin{equation}
\begin{split}
A(z)&=\frac{1}{W(n)}\Wrons\left({\vs{\Phi}}(n;z),\vs{\Psi}(n;z)\right),\\
B(z)&=\frac{1}{W(n)}\Wrons\left({\ovl{\vs{\Psi}}(n;z),\vs{\Phi}}(n;z)\right),
\end{split}
\end{equation}
where
\begin{equation}
W(n)=\Wrons\left(\ovl{\vs{\Psi}}(n;z),\vs{\Psi}(n;z)\right).
\end{equation}
For the implicit Euler method, we have the recurrence relation
\begin{equation}
W(n+1)=\frac{1}{\Theta(n+1)}W(n),
\end{equation}
while $W(n)=1$ for the split-Magnus method. For the trapezoidal rule, we have
\begin{equation}
W(n+1)=\frac{\Theta(n)}{\Theta(n+1)}W(n).
\end{equation}

Let us define the matrix Jost solution by 
\begin{equation}
\begin{split}
v(n;z) 
&= \left(\vs{\Phi}(n;z),\vs{\Psi}(n;z)\right)\\
&=\begin{pmatrix}
\Phi_1(n;z) & \Psi_1(n;z)\\
\Phi_2(n;z) & \Psi_2(n;z)
\end{pmatrix},
\end{split}
\end{equation}
so that
\begin{equation}
\begin{split}
{v}(n+1;z)&=
\frac{z^{-1}}{\Delta(n)}
\begin{pmatrix}
1 & S(n)\\
T(n)& 1
\end{pmatrix}\Lambda(z^2)\vv{v}(n;z)\\
&=z^{-1}M(n+1;z^2){v}(n;z),
\end{split}
\end{equation}
where $\Lambda(z)=\diag(1,z)$. Let $\mathfrak{S}_K$ denote the discrete spectrum to 
be ``added'' to the seed potential $S_0(n)$ so that the augmented potential 
is denoted by $S_K(n)$. Guided by the $z^2$ dependence above, we may introduce 
the Darboux matrix
\begin{equation*}
D_{K}(n;z^2,\mathfrak{S}_K)=\sum_{k=-K}^{K}D_k^{(K)}(n;\mathfrak{S}_K)z^{2k},
\end{equation*}
such that
\begin{equation*}
{v}_K(n;z)=D_{K}(n;z^2,\mathfrak{S}_K)v_0(n;z)\Gamma_{K}(z^2),
\end{equation*}
where $\Gamma_K=\diag(\gamma_1^{(K)},\gamma_2^{(K)})$ is
introduced in order to correct for the scaling factors in the asymptotes as
$n\rightarrow\pm\infty$. The compatibility relation between the transfer matrix
and the Darboux matrix reads as
\begin{equation}\label{eq:DT-compat-bdf1}
D_{K}(n+1;z^2)M_0(n+1;z^2)=M_K(n+1;z^2)D_{K}(n;z^2),
\end{equation}
where we have suppressed the dependence on $\mathfrak{S}_K$ for the sake of
brevity. From the general symmetry property of the transfer matrix, 
$z^{-1}M(n+1;z^2)=z\ovl{M}(n+1;z^2)$, it follows that
\begin{equation}
\ovl{D}_{K}(n;z^2)
=\sum_{k=-K}^{K}
(i\sigma_2)D_k^{(K)*}(n)(i\sigma_2)^{\dagger}z^{-2k}={D}_{K}(n;z^2);
\end{equation}
therefore,
\begin{equation}
D_{-k}^{(K)}(n)=\sigma_2D_k^{(K)*}(n)\sigma_2.
\end{equation}
For $k=0$, this translates into the requirement that
\begin{equation}
D^{(K)}_0(n)
=
\begin{pmatrix}
d^{(K,0)}_{1}(n)& d^{(K,0)}_{2}(n)\\
-d^{(K,0)*}_{2}(n)& d^{(K,0)*}_{1}(n)
\end{pmatrix},\\
\end{equation}
where $d^{(K,0)}_{1}(n)$ and $d^{(K,0)}_{2}(n)$ are to be determined. Now,
coming back to the compatibility relation~\eqref{eq:DT-compat-bdf1} and 
equating the coefficient of $z^{2K+2}$ on both sides, we have
\begin{multline*}
\begin{pmatrix}
d^{(K,K)}_{11}(n+1)& d^{(K,K)}_{12}(n+1)\\
d^{(K,K)}_{21}(n+1)& d^{(K,K)}_{22}(n+1)
\end{pmatrix}
\begin{pmatrix}
0& S_0(n)\\
0& 1
\end{pmatrix}\\
=\frac{\Delta_0(n)}{\Delta_K(n)}
\begin{pmatrix}
0& S_K(n)\\
0& 1
\end{pmatrix}
\begin{pmatrix}
d^{(K,K)}_{11}(n)& d^{(K,K)}_{12}(n)\\
d^{(K,K)}_{21}(n)& d^{(K,K)}_{22}(n)
\end{pmatrix}.
\end{multline*}
It follows that $d^{(K,K)}_{21}(n)=0$ and
\begin{equation}
d^{(K,K)}_{22}(n+1)
=\frac{\Delta_0(n)}{\Delta_K(n)}d^{(K,K)}_{22}(n).
\end{equation}
Making the choice $d^{(K,K)}_{11}(n)=0$ allows us to conclude
\begin{equation}
d^{(K,K)}_{12}(n+1)
=\frac{\Delta_0(n)}{\Delta_K(n)}S_K(n)d^{(K,K)}_{22}(n),
\end{equation}
which yields the first important identity for the discrete DT:
\begin{equation}
S_K(n-1)=\frac{d^{(K,K)}_{12}(n)}{d^{(K,K)}_{22}(n)},\quad n\in\field{Z}.
\end{equation}
Equating the coefficient of $z^{2K}$, we have
\begin{equation}
T_0(n)d^{(K,K)}_{22}(n+1)
=\frac{\Delta_0(n)}{\Delta_K(n)}d^{(K,K-1)}_{21}(n),
\end{equation}
which translates into 
\begin{equation}
d^{(K,K-1)}_{21}(n)=
T_0(n)d^{(K,K)}_{22}(n).
\end{equation}
From this point onwards, we discuss each of the discrete systems separately.

\subsection{Implicit Euler method}\label{sec:DDT-IE}
For the implicit Euler method, we have $\Delta(n)=\Theta(n+1)$ and
$S(n)=Q(n+1)$ so that
\begin{equation}
\begin{split}
&d^{(K,K)}_{11}(n)=0,\quad d^{(K,K)}_{21}(n)=0,\\
&d^{(K,K)}_{12}(n+1)=\frac{\Theta_0(n+1)}{\Theta_K(n+1)}Q_K(n+1)d^{(K,K)}_{22}(n),\\
&d^{(K,K)}_{22}(n+1)=\frac{\Theta_0(n+1)}{\Theta_K(n+1)}d^{(K,K)}_{22}(n),
\end{split}
\end{equation}
which yields
\begin{equation}
Q_K(n)=\frac{d^{(K,K)}_{12}(n)}{d^{(K,K)}_{22}(n)},\quad n\in\field{Z}.
\end{equation}
Further,
\begin{equation}
d^{(K,K-1)}_{21}(n)=
R_0(n+1)d^{(K,K)}_{22}(n).
\end{equation}
From this point onwards, we consider the particular case of $K=1$. It turns out
that in this case, it is possible to obtain explicit expressions for the entries
of the Darboux matrix. It is clear from the discussion above that the Darboux
matrix coefficients can chosen such that
\begin{equation}
D_{1}^{(1)}(n)=
\begin{pmatrix}
0&  d^{(1,1)}_{1}(n)\\
0&  d^{(1,1)}_{2}(n)
\end{pmatrix},\quad D_{-1}^{(1)}(n)=
\begin{pmatrix}
d^{(1,1)*}_{2}(n)& 0\\
-d^{(1,1)*}_{1}(n)& 0
\end{pmatrix}.
\end{equation}
Next, from the definition of the norming constant $b_1$, we have
\begin{equation}\label{eq:linear-system-bdf1}
D_{1}(n;z^2_1)
[\vs{\Phi}_0(n;z_1)-B_{1}\vs{\Psi}_0(n;z_1)]=0,
\end{equation}
where $B_1=b_1\gamma^{(1)}_2(z^2_1)/\gamma^{(1)}_1(z^2_1)$. This approach is
entirely similar to that of Neugebauer and Meinel~\cite{NM1984} for the
continuous case. Introducing 
\begin{equation}
\beta_0(n;z_1)=\frac{{\Phi}^{(0)}_1(n;z_1)-B_{1}{\Psi}^{(0)}_1(n;z_1)}
{{\Phi}^{(0)}_2(n;z_1)-B_{1}{\Psi}^{(0)}_2(n;z_1)}
\end{equation}
the linear system in~\eqref{eq:linear-system-bdf1} reads as
\begin{widetext}
\begin{equation}
\begin{split}
&d^{(1,0)}_1(n)\beta_0(n;z_1)+d^{(1,0)}_2(n)
=-d^{(1,1)}_1(n)z_1^2-d^{(1,1)*}_2(n)\frac{\beta_0(n;z_1)}{z_1^{2}},\\
&d^{(1,0)}_1(n)-\beta^*_0(n;z_1)d^{(1,0)}_2(n)
=d^{(1,1)}_1(n)\frac{\beta^*_0(n;z_1)}{z^{*2}_1}-d^{(1,1)*}_2(n)z_1^{*2}.
\end{split}
\end{equation}
Solving this linear system for $d^{(1,0)}_1(n)$ and $d^{(1,0)}_2(n)$, yields
\begin{equation}\label{eq:soln-bdf1}
\begin{split}
d_1^{(1,0)}(n)&=-\frac{1}{1+|\beta_0(n;z_1)|^2}
\left[d_1^{(1,1)}(n)\left(z_1^2-\frac{1}{z^{*2}_1}\right)\beta^*_0(n;z_1)
+d_2^{(1,1)*}(n)\Xi_1^*(n)\right],\\
d_2^{(1,0)}(n)&=
-\frac{1}{1+|\beta_0(n;z_1)|^2}
\left[d_1^{(1,1)}(n)\Xi_1(n)+d_2^{(1,1)*}(n)
\left(\frac{1}{z_1^2}-z^{*2}_1\right)\beta_0(n;z_1)\right],
\end{split}
\end{equation}
where
\begin{equation}\label{eq:Xi-def}
\Xi_1(n)={\left(z_1^{2}+\frac{|\beta_0(n;z_1)|^2}{z^{*2}_1}\right)}.
\end{equation}
Now, from $d^{(1,0)}_{2}(n)=Q_0(n+1)d^{(1,1)*}_{2}(n)$, it follows that
\begin{equation}
{\Xi_1(n)}\frac{d_1^{(1,1)}(n)}{d_2^{(1,1)*}(n)}
=-\left(1+|\beta_0(n;z_1)|^2\right)Q_0(n+1)\\
-\left(\frac{1}{z_1^2}-z^{*2}_1\right)\beta_0(n;z_1).
\end{equation}
\end{widetext}
From the expressions above, it is clear that $|d_2^{(1,1)}(n)|$ is a free scale parameter
so that it can be set to unity. However, the phase of $d_2^{(1,1)}(n)$ is not arbitrary.
We find that the choice $d_2^{(1,1)}(n)=i$ conforms with the limit to continuum 
so that $d^{(1,1)}_{1}(n)=iQ_1(n)$. Consequently, the augmented
potential is given by
\begin{equation}
Q_1(n)=\frac{1+|\beta_0(n;z_1)|^2}{\Xi_1(n)}Q_0(n+1)
+\left(\frac{1}{z_1^2}-z^{*2}_1\right)\frac{\beta_0(n;z_1)}{\Xi_1(n)}.
\end{equation}
While this expression bears a close resemblance to that of the continuous case,
there is one striking difference: The augmented potential at the grid point
$t_n$ requires the seed potential at the grid point $t_{n+1}$. Given that, in
practice, we restrict ourselves to a finite grid, the knowledge of the potential
on the edge must be provided or assumed to be zero. Therefore, in order to avoid boundary
effects, one can use the continuous DT at $t_0=0$ to compute the potential at all
points to the left of origin using the discrete DT. Symmetry properties of the ZS problem 
can then be used to compute the potential to the right of the origin by repeating this procedure
with $1/B_1$ instead of $B_1$. Finally, let us summarize the entries of the
Darboux matrix in a form that is more suited for implementation in a computer
program:
\begin{multline}
d_1^{(1,0)}(n)=
-iQ_0(n+1)\left(z_1^2-\frac{1}{z^{*2}_1}\right)\frac{\beta^*_0(n;z_1)}{\Xi_1(n)}\\
+\frac{i}{\Xi_1(n)}{\left(|z_1|^{4}+\frac{|\beta_0(n;z_1)|^2}{|z_1|^4}\right)},
\end{multline}
and,
\begin{multline}
d^{(1,1)}_{1}(n)=iQ_0(n+1)\frac{1+|\beta_0(n;z_1)|^2}{\Xi_1(n)}\\
+\left(\frac{1}{z_1^2}-z^{*2}_1\right)\frac{i\beta_0(n;z_1)}{\Xi_1(n)}
\end{multline}
together with $d_2^{(1,0)}(n)=-iQ_0(n+1)$ and $d_2^{(1,1)}=i$ so that $Q_1(n)=-id^{(1,1)}_{1}(n)$.

In order to see how the scattering data changes as a result of addition of one
soliton, we write the Darboux matrix in the form
\begin{multline}
D_1(n;z^2) = 
\begin{pmatrix}
d^{(1,1)*}_{2}(n) & d^{(1,1)}_{1}(n)\\
-d^{(1,1)*}_{1}(n)& d^{(1,1)}_{2}(n)
\end{pmatrix}\\
\times\begin{pmatrix}
-\beta_0(n;z_1)z^{-2}z_1^{-2}
& 1\\
1            
& \beta^*_0(n;z_1)z^2z^{*-2}_1
\end{pmatrix}
\begin{pmatrix}
z^2-z_1^{2} & 0\\
0              &\frac{1}{z^2}-z_1^{*2} \\
\end{pmatrix}\times\\
\frac{1}{1+|\beta_0(n;z_1)|^2}\begin{pmatrix}
\beta^*_0(n;z_1) & 1\\
1              & -\beta_0(n;z_1)\\
\end{pmatrix}.
\end{multline}
From here, it is straightforward to conclude that
\begin{multline}
\det[D_1(n;z^2)]=(z^2-z_1^{2})\left(\frac{1}{z^2}-z_1^{*2}\right)\times\\
\left(\frac{|z_1|^4+|\beta_0(n;z_1)|^2}{1+|\beta_0(n;z_1)|^2}\right)
\frac{\Theta_1(n)}{|z_1|^4}.
\end{multline}
Now, the asymptotic form of the Darboux matrix as $n\rightarrow-\infty$
works out to be
\begin{equation}
D_{1}(n;z^2)
\rightarrow
\begin{pmatrix}
-i\left(\frac{1}{z^2}-z_1^{*2}\right) & 0\\
0 & i(z^2-z_1^2)
\end{pmatrix},
\end{equation}
which allows us to conclude
\begin{equation}
\gamma^{(1)}_1(z^2)=\frac{i}{1/z^2 - z^{*2}_1}.
\end{equation}
Similarly, as $n\rightarrow+\infty$, we have
\begin{equation}
D_{1}(n;z^2,\mathfrak{S}_1)
\rightarrow
\begin{pmatrix}
-i(\frac{1}{z^2}-\frac{1}{z_1^{2}}) & 0\\
0 & i(z^2-\frac{1}{z_1^{*2}})
\end{pmatrix},
\end{equation}
which allows us to conclude
\begin{equation}
\gamma^{(1)}_2(z^2)=\frac{-i}{z^2-{1}/{z_1^{*2}}}
=\frac{i(z_1^{*}/z)^2}{1/z^2 - z^{*2}_1}.
\end{equation}
Once the scale factor is determined, the $A$ coefficient works out to be 
\begin{equation}
\begin{split}
A_1(z^2) &= \frac{1}{W_1(n)}\Wrons\left(\vs{\Phi}_1(n;z),\vs{\Psi}_1(n;z)\right)\\
&=\frac{W_0(n)}{W_1(n)}\det[D_{1}(n;z^2)]A_0(z^2)\det[\Gamma_1(z^2)]\\
&=\left(\frac{1/z^2 -1/z^{2}_1}{1/z^2 -z^{*2}_1}\right) A_0(z^2),
\end{split}
\end{equation}
where
\begin{equation}
W_1(n)=\left(\frac{|z_1|^4+|\beta_0(n;z_1)|^2}{1+|\beta_0(n;z_1)|^2}\right)
\Theta_1(n)W_0(n).
\end{equation}
Finally, the discrete version of the norming constant is given by
\begin{equation}
B_1 = b_1\left(\frac{z_1^*}{z_1}\right)^2.
\end{equation}
A straightforward application of the method developed in this section yields the 
explicit form of the discrete one soliton solution:
\begin{equation}
\begin{split}
Q(n)&=-\left(\frac{1}{z_1^2}-z^{*2}_1\right)\frac{(1/B_1)}
{(z_1^2)^{n+1}+\frac{1}{|B_1|^2}(1/z^{*2})^{n+1}}\\
&=-\frac{\sinh(2\eta_1 h)}
{\cosh[2\eta_1(n+1)h-\kappa_1]}e^{-2i\xi_1nh-i\theta_1},
\end{split}
\end{equation}
where $b_1=e^{\kappa_1+i\theta_1}$.
Therefore,
\begin{equation}
\frac{1}{h}Q(n)
=-\frac{2\eta_1}{\cosh[2\eta_1t_n-\kappa_1]}
e^{-2i\xi_1t_n-i\theta_1}+\bigO{h}.
\end{equation}

\subsection{Split-Magnus Method}
Let us recall that for the split-Magnus case, we have $\vv{v}(n) =
\Lambda(z)\vv{w}(n)$ and
\begin{equation}
\vv{w}(n+1)=\frac{z^{-1}}{\breve{\Theta}^{1/2}(n+1)}
\begin{pmatrix}
1& \breve{Q}(n+1)\\
\breve{R}(n+1)& 1
\end{pmatrix}\Lambda(z^2)\vv{w}(n),
\end{equation}
where $\breve{Q}(n)=Q(n-{1}/{2})$ and the same convention holds for
$\breve{R}(n)$ and $\breve{\Theta}(n)$. In terms of the matrix Jost solution, the 
discrete DT reads as
\begin{equation*}
{v}_K(n;z)=\Lambda(z)D_{K}(n;z^2,\mathfrak{S}_K)\Lambda(z^{-1})v_0(n;z)\Gamma_{K}(z^2),
\end{equation*}
where $\Gamma_K$ and $D_{K}(n;z^2,\mathfrak{S}_K)$ are as defined in the last
section. By analogy, the following relations follow quite trivially:
\begin{equation*}
\begin{split}
&d^{(K,K)}_{21}(n)=0,\quad d^{(K,K)}_{11}(n)=0\\
&d^{(K,K)}_{22}(n+1)
=\frac{\breve{\Theta}^{1/2}_0(n+1)}{\breve{\Theta}^{1/2}_K(n+1)}d^{(K,K)}_{22}(n),\\
&d^{(K,K)}_{12}(n+1)
=\frac{\breve{\Theta}^{1/2}_0(n+1)}{\breve{\Theta}^{1/2}_K(n+1)}\breve{Q}_K(n+1)d^{(K,K)}_{22}(n),
\end{split}
\end{equation*}
which yields
\begin{equation}
\breve{Q}_K(n+1)=\frac{d^{(K,K)}_{12}(n+1)}{d^{(K,K)}_{22}(n+1)},\quad n\in\field{Z}.
\end{equation}
Further, we have
\begin{equation}
d^{(K,K-1)}_{21}(n)=\breve{R}_0(n+1)d^{(K,K)}_{22}(n).
\end{equation}
Again, we specialize to the case $K=1$. Introducing the matrix $D_1(n;z^2)$ as in 
the case of implicit Euler method, the
relation~\eqref{eq:linear-system-bdf1} gets modified to
\begin{equation}\label{eq:linear-system-sm}
D_{1}(n;z^2_1)\Lambda(z_1^{-1})
[\vs{\Phi}_0(n;z_1)-B_{1}\vs{\Psi}_0(n;z_1)]=0.
\end{equation}
Introducing
\begin{equation}
\beta_0(n;z_1)=z_1\frac{{\Phi}^{(0)}_1(n;z_1)-B_{1}{\Psi}^{(0)}_1(n;z_1)}
{{\Phi}^{(0)}_2(n;z_1)-B_{1}{\Psi}^{(0)}_2(n;z_1)},
\end{equation}
we note that the solution of the linear system~\eqref{eq:linear-system-sm} is
also given by~\eqref{eq:soln-bdf1} which specifies $d_1^{(1,0)}(n)$ and 
$d_2^{(1,0)}(n)$ in terms of $d_1^{(1,1)}(n)$ and $d_2^{(1,1)}(n)$. Further, 
the variable $|d_2^{(1,1)}(n)|$ was identified as a free scale parameter in the
previous case; however, in the present case it can no
longer be chosen arbitrarily on account of the condition
\begin{equation}
\det[D_1(n+1;z^2)] = \det[D_1(n;z^2)],
\end{equation}
which follows from the compatibility relation~\eqref{eq:DT-compat-bdf1}. Choosing 
the phase to be the same as before, let 
$d_2^{(1,1)}(n)=i\alpha(n)$ so that $d^{(1,1)}_{1}(n)=i\breve{Q}_1(n)\alpha(n)$.
The augmented potential then works out to be
\begin{equation}
\breve{Q}_1(n)=\frac{1+|\beta_0(n;z_1)|^2}{\Xi_1(n)}\breve{Q}_0(n+1)
+\left(\frac{1}{z_1^2}-z^{*2}_1\right)\frac{\beta_0(n;z_1)}{\Xi_1(n)}.
\end{equation}
The local error in the expression about with respect to $h$ can be obtained by
a Taylor expansion. Observing from~\eqref{eq:Xi-def},
\begin{equation}
\begin{split}
\frac{1+|\beta_0(n;z_1)|^2}{\Xi_1(n)} 
=\frac{1+|\beta_0(n;z_1)|^2}{\left(z_1^{2}+\frac{|\beta_0(n;z_1)|^2}{z^{*2}_1}\right)}
=1+\bigO{h},
\end{split}
\end{equation}
it is follows that the second order of accuracy cannot propagate to the next level
unless $\breve{Q}_0(n+1)$ is identically zero. This is only true of the null
potential; therefore, despite the fact that the underlying discrete framework has
an accuracy of $\bigO{h^2}$, the DT iterations for multisolitons has first
order accuracy beyond the one-soliton case, which as we will see below turns out to
be accurate to $\bigO{h^2}$ .

Now, in order to determine $\alpha(n)$, we consider the determinant of the Darboux
matrix which is given by
\begin{multline}
\det[D_1(n;z^2)]=(z^2-z_1^{2})\left(\frac{1}{z^2}-z_1^{*2}\right)\times\\
\alpha^2(n)\left(\frac{|z_1|^4+|\beta_0(n;z_1)|^2}{1+|\beta_0(n;z_1)|^2}\right)
\frac{\breve{\Theta}_1(n)}{|z_1|^4}.
\end{multline}
It can be inferred from the compatibility relation~\eqref{eq:DT-compat-bdf1} 
between the Darboux matrix and the transfer matrix that the determinant of the 
Darboux matrix must be independent of $n$ because the same is true of the transfer 
matrix. Therefore, we choose
\begin{equation}
\alpha(n)=
\frac{|z_1|^2}{\breve{\Theta}^{1/2}_1(n)}
\left(\frac{1+|\beta_0(n;z_1)|^2}{|z_1|^4+|\beta_0(n;z_1)|^2}\right)^{1/2},
\end{equation}
so that
\begin{equation}
\det[D_1(n;z^2)]=(z^2-z_1^{2})\left(\frac{1}{z^2}-z_1^{*2}\right).
\end{equation}
From the asymptotic forms, we can also conclude
\begin{equation}
\begin{split}
\gamma^{(1)}_1(z^2)&=\frac{i}{1/z^2 - z^{*2}_1},\\
\gamma^{(1)}_2(z^2)&=\frac{iz^{-2}}{1/z^2 - z^{*2}_1}
\left(\frac{z_1^{*}}{z_1}\right),
\end{split}
\end{equation}
so that the $A$ coefficient works out to be
\begin{equation}
\begin{split}
A_1(z^2) &= \Wrons\left(\vs{\Phi}_1(n;z),\vs{\Psi}_1(n;z)\right)\\
&=\det[D_{1}(n;z^2)]A_0(z^2)\det[\Gamma_1(z^2)]\\
&=\left(\frac{|z_1|^2}{z_1^2}\cdot\frac{z^2 - z^{2}_1}{z^2z^{*2}_1-1}\right) A_0(z^2),
\end{split}
\end{equation}
where the expression in the parenthesis is the well-known Blaschke product. Finally, 
the discrete version of the norming constant is given by
\begin{equation}
B_1 = b_1\left(\frac{z_1^*}{z^3_1}\right).
\end{equation}

We conclude this discussion with the one soliton solution:
\begin{equation}
\begin{split}
Q(n-\tfrac{1}{2})&=-\left(\frac{1}{z_1^2}-z^{*2}_1\right)
\frac{(1/b_1)(z_1/z_1^*)}{(z_1)^{2n-1}+\frac{1}{|b_1|^2}(1/z^{*})^{2n-1}}\\
&=-\frac{\sinh(2\eta_1 h)}
{\cosh[2\eta_1(n-\tfrac{1}{2})h-\kappa_1]}e^{-2i\xi_1(n-\frac{1}{2})h-i\theta_1}.
\end{split}
\end{equation}
Therefore,
\begin{equation}
\frac{1}{h}Q_1(n-\tfrac{1}{2})
=-\frac{2\eta_1}{\cosh[2\eta_1t_{n-1/2}-\kappa_1]}
e^{-2i\xi_1t_{n-1/2}-i\theta_1}+\bigO{h^2}.
\end{equation}
As discussed earlier, the second order of convergence seen here only holds for
one soliton potential.

\subsection{Trapezoidal rule}
Based on the transfer matrix relation~\eqref{eq:TM-TR}, the discrete DT in terms 
of the matrix Jost solution, reads as
\begin{multline}
{v}_K(n;z)=
\frac{1}{\Theta_K(n)}
\begin{pmatrix}
1        & -Q_K(n)\\
-R_K(n) & 1
\end{pmatrix}D_{K}(n;z^2,\mathfrak{S}_K)\times\\
\begin{pmatrix}
1        & Q_0(n)\\
R_0(n) & 1
\end{pmatrix}v_0(n;z)\Gamma_{K}(z^2),
\end{multline}
where $\Gamma_K$ and $D_{K}(n;z^2,\mathfrak{S}_K)$ are as defined in 
Sec.~\ref{sec:DDT-IE}. 
Recalling that $\Delta(n)={\Theta(n+1)}/{[2-\Theta(n+1)]}$, we have
\begin{equation*}
\begin{split}
&d^{(K,K)}_{21}(n)=0,\quad d^{(K,K)}_{11}(n)=0\\
&d^{(K,K)}_{22}(n+1)
=\frac{[2-\Theta_K(n+1)]}{[2-\Theta_0(n+1)]}
\frac{\Theta_0(n+1)}{\Theta_K(n+1)}d^{(K,K)}_{22}(n),\\
&d^{(K,K)}_{12}(n+1)
=\frac{[2-\Theta_K(n+1)]}{[2-\Theta_0(n+1)]}
\frac{\Theta_0(n+1)}{\Theta_K(n+1)}
\breve{Q}_K(n+1)d^{(K,K)}_{22}(n),
\end{split}
\end{equation*}
which yields
\begin{equation}
{Q}_K(n+1)=\frac{d^{(K,K)}_{12}(n+1)}{d^{(K,K)}_{22}(n+1)},\quad n\in\field{Z}.
\end{equation}
Further, we have
\begin{equation}
d^{(K,K-1)}_{21}(n)={R}_0(n+1)d^{(K,K)}_{22}(n).
\end{equation}
Once again, specializing to the case $K=1$, let the matrix $D_1(n;z^2)$ be
defined as in the case of the implicit Euler method so that the 
relation~\eqref{eq:linear-system-bdf1} gets modified to
\begin{equation}\label{eq:linear-system-tr}
D_{1}(n;z^2_1)
\begin{pmatrix}
1        & Q_0(n)\\
R_0(n) & 1
\end{pmatrix}[\vs{\Phi}_0(n;z_1)-B_{1}\vs{\Psi}_0(n;z_1)]=0.
\end{equation}
Introducing
\begin{equation}
\beta'_0(n;z_1)=\frac{{\Phi}^{(0)}_1(n;z_1)-B_{1}{\Psi}^{(0)}_1(n;z_1)}
{{\Phi}^{(0)}_2(n;z_1)-B_{1}{\Psi}^{(0)}_2(n;z_1)}
\end{equation}
and
\begin{equation}
\beta_0(n;z_1)=\frac{\beta'_0(n;z_1)+Q_0(n)}{\beta'_0(n;z_1)R_0(n)+1},
\end{equation}
we note that the solution of the linear system~\eqref{eq:linear-system-tr} is
also given by~\eqref{eq:soln-bdf1} which specifies $d_1^{(1,0)}(n)$ and 
$d_2^{(1,0)}(n)$ in terms of $d_1^{(1,1)}(n)$ and $d_2^{(1,1)}(n)$. Again, from the 
compatibility relation~\eqref{eq:DT-compat-bdf1}, it follows that
\begin{equation}
\det[D_1(n+1;z^2)] = \det[D_1(n;z^2)].
\end{equation}
In order satisfy this condition, we set $d_2^{(1,1)}(n)=i\alpha(n)$ 
so that $d^{(1,1)}_{1}(n)=iG_1(n)\alpha(n)$ which in turn leads to the expression 
for the augmented potential
\begin{equation}\label{eq:DT-TR-G1-G0}
G_1(n)=\frac{1+|\beta_0(n;z_1)|^2}{\Xi_1(n)}G_0(n+1)
+\left(\frac{1}{z_1^2}-z^{*2}_1\right)\frac{\beta_0(n;z_1)}{\Xi_1(n)}.
\end{equation}
The actual potential $Q_1(n)$ can be determined from $G_1(n)$ as 
\begin{equation}
Q_1(n)=\frac{G_1(n)}{1+\sqrt{1+|G_1(n)|^2}}.
\end{equation}
The choice of
\begin{equation}
\alpha(n)=
\frac{[2-\Theta_1(n)]|z_1|^2}{\Theta_1(n)}
\left(\frac{1+|\beta_0(n;z_1)|^2}{|z_1|^4+|\beta_0(n;z_1)|^2}\right)^{1/2},
\end{equation}
leads to
\begin{equation}
\begin{split}
\gamma^{(1)}_1(z^2)&=\frac{i}{1/z^2 - z^{*2}_1},\\
\gamma^{(1)}_2(z^2)&=\frac{iz^{-2}}{1/z^2 -
z^{*2}_1}\left(\frac{z_1^{*}}{z_1}\right),
\end{split}
\end{equation}
so that
\begin{equation}
B_1 = b_1\left(\frac{z_1^*}{z^3_1}\right).
\end{equation}
Again, on account of the $\bigO{h}$ contribution from the first term in the
right hand side of~\eqref{eq:DT-TR-G1-G0}, the DT iteration for the trapezoidal
rule can never achieve second order of convergence unless the seed solution
corresponds to the null potential. 

We conclude this discussion with the one soliton solution:
\begin{equation}
\begin{split}
G_1(n)&=-\left(\frac{1}{z_1^2}-z^{*2}_1\right)
\frac{(1/b_1)(z_1/z_1^*)}{(z_1)^{2n}+\frac{1}{|b_1|^2}(1/z^{*})^{2n}}\\
&=-\frac{\sinh(2\eta_1 h)}
{\cosh[2\eta_1nh-\kappa_1]}e^{-2i\xi_1nh-i\theta_1}.
\end{split}
\end{equation}
Therefore,
\begin{equation}
\frac{1}{h}Q_1(n)
=-\frac{2\eta_1}{\cosh[2\eta_1t_{n}-\kappa_1]}
e^{-2i\xi_1t_{n}-i\theta_1}+\bigO{h^2}.
\end{equation}

\begin{figure}[!h]
\centering
\includegraphics[scale=1]{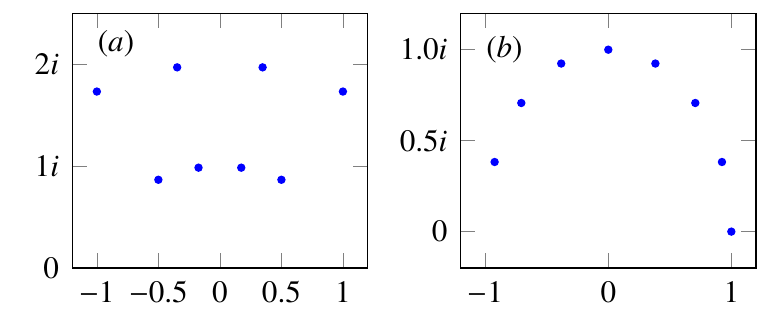}
\caption{\label{fig:spec}The figure shows the eigenvalues and the norming
constant for an $8$-soliton solutions.}
\end{figure}

\begin{figure*}[!t]
\centering
\includegraphics[scale=1]{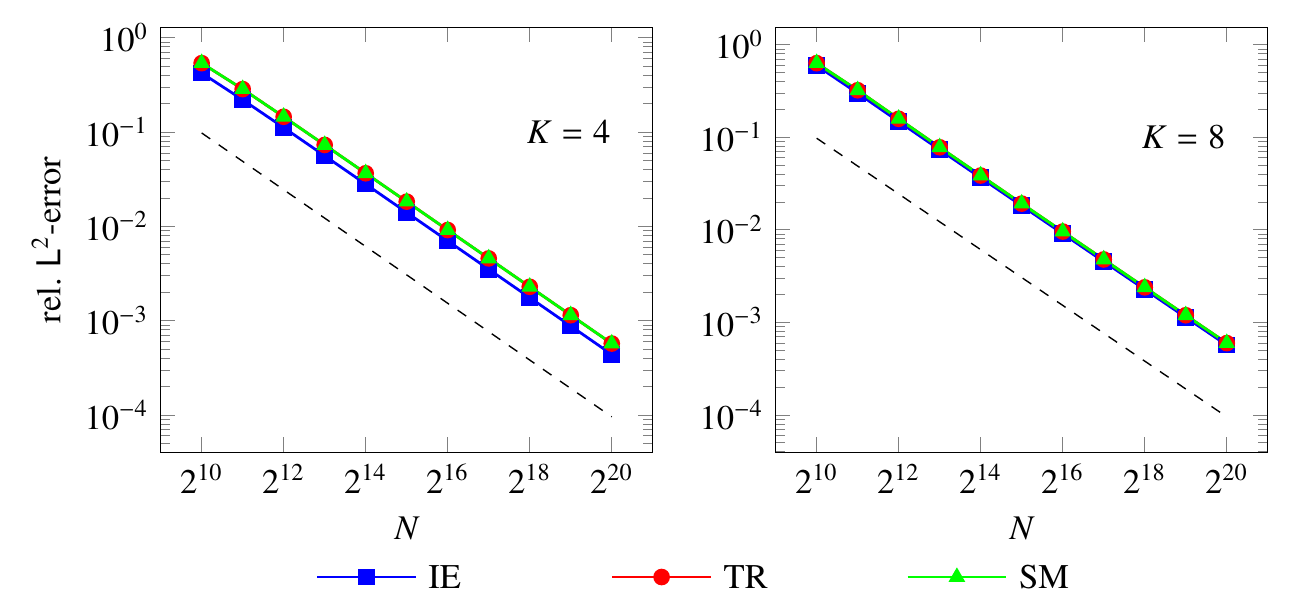}
\caption{\label{fig:convg} The figure shows the convergence analysis for the
discrete Darboux transformation for $K$-soliton solutions ($K=4,8$) 
corresponding to the discrete spectrum defined in 
Sec.~\ref{sec:num-test} with respect to the number of samples $N$. The discrete
systems correspond to the implicit Euler (IE) method, the trapezoidal rule (TR) and the
split-Magnus (SM) method. The dashed line depicts the $\bigO{N^{-1}}$ curve for reference.}
\end{figure*}
\section{Numerical Test and Conclusion}\label{sec:num-test}
A simple numerical test can be designed to confirm the rate of convergence of
the discrete Darboux transformation derived in the earlier sections. Define
\begin{equation}
\vs{\theta} =\left(\frac{\pi}{3}, \frac{13\pi}{30}, 
                   \frac{8\pi}{15}, \frac{19\pi}{30}\right),
\end{equation}
and let the set of eigenvalues be $\{\exp(i\vs{\theta}),2\exp(i\vs{\theta})\}$.
The corresponding norming constants are chosen as
\begin{equation}
b_j=\exp{[i(\pi/8)(j-1)]},\quad j=1,2,\ldots,8.
\end{equation}
The discrete spectrum defined above is shown in Fig.~\ref{fig:spec}.
Let the time domain be $\Omega=[-10,10]$ and the grid be defined as
$t_n=nh,\,n=-N/2,\ldots,N/2-1,$
where $h = 20/N$. The number of samples $N$ varies within the set 
$\{2^{10},2^{11},\ldots,2^{20}\}$. The error in computing the $K$-soliton
solutions is quantified by
\begin{equation}
e_{\text{rel.}}=\|q-q^{\text{num.}}\|_{\fs{L}^2(\Omega)}/\|q\|_{\fs{L}^2(\Omega)},
\end{equation}
where the integrals are estimated using the trapezoidal rule. The exact solution
is computed using the classical Darboux transformation. The convergence
analysis for the $4$-soliton and the $8$-soliton solutions are shown in
Fig.~\ref{fig:convg} which clearly indicates that rate of convergence is
$\bigO{N^{-1}}$ for each of the methods considered.

\appendix
%

\providecommand{\noopsort}[1]{}\providecommand{\singleletter}[1]{#1}%

\end{document}